\documentclass[sigconf]{acmart}
\AtBeginDocument{%
  \providecommand\BibTeX{{%
    \normalfont B\kern-0.5em{\scshape i\kern-0.25em b}\kern-0.8em\TeX}}}

\setcopyright{acmcopyright}
\copyrightyear{2021}
\acmYear{2021}

\acmConference[KDD 2021]{Woodstock '18: ACM Symposium on Neural
  Gaze Detection}{August 14--18, 2021}{Singapore}
\acmBooktitle{KDD}

\usepackage{multirow}
\usepackage{subfigure}

\begin{document}
\fancyhead{}
\title{Exploring Self-Supervised Representation Ensembles for COVID-19 Cough Classification}
\author{Hao Xue}
\email{hao.xue@rmit.edu.au}
\orcid{0000-0003-1700-9215}
\affiliation{%
  \institution{School of Computing Technologies, RMIT University}
  \city{Melbourne}
  \state{VIC}
  \country{Australia}
}

\author{Flora D. Salim}
\email{flora.salim@rmit.edu.au}
\orcid{0000-0002-1237-1664}
\affiliation{%
  \institution{School of Computing Technologies, RMIT University}
  \city{Melbourne}
  \state{VIC}
  \country{Australia}
}
\renewcommand{\shortauthors}{Xue and Salim}

\begin{abstract}
The usage of smartphone-collected respiratory sound, trained with deep learning models, for detecting and classifying COVID-19 becomes popular recently. It removes the need for in-person testing procedures especially for rural regions where related medical supplies, experienced workers, and equipment are limited. However, existing sound-based diagnostic approaches are trained in a fully-supervised manner, which requires large scale well-labelled data. It is critical to discover new methods to leverage unlabelled respiratory data, which can be obtained more easily. In this paper, we propose a novel self-supervised learning enabled framework for COVID-19 cough classification. A contrastive pre-training phase is introduced to train a Transformer-based feature encoder with unlabelled data. Specifically, we design a random masking mechanism to learn robust representations of respiratory sounds. The pre-trained feature encoder is then fine-tuned in the downstream phase to perform cough classification. In addition, different ensembles with varied random masking rates are also explored in the downstream phase. Through extensive evaluations, we demonstrate that the proposed contrastive pre-training, the random masking mechanism, and the ensemble architecture contribute to improving cough classification performance.
\end{abstract}



\maketitle
\newcommand{\eg}{\textit{e.g.}}
\newcommand{\etal}{\textit{et al.}}

\section{Introduction}

By February 1\textsuperscript{st} 2021, the total number of coronavirus disease 2019 (COVID-19) confirmed cases exceeded 103 million world-wide\footnote{https://en.wikipedia.org/wiki/COVID-19\_pandemic\_by\_country\_and\_territory}, and at the time of writing, it is still an ongoing pandemic.
Given that the global vaccination effort is still in its early stage, a practical and effective defensive procedure against the highly contagious COVID-19 is large-scale and timely testing, aimed at detecting and isolating the infected individuals as soon as possible.
Developing a reliable, easily-accessible, and contactless approach for preliminary diagnosis of COVID-19 is significant.
It will also benefit regions where medical supplies/workers and personal protective equipment are limited.
As pointed out by Imran~\etal~\cite{imran2020ai4covid}, cough is one of the major symptoms of COVID-19 patients. 
Compared to PCR (Polymerase Chain Reaction) tests and radiological images, diagnosis using cough sounds can be easily accessed by people through a smartphone app.
In the meantime, however, cough is also a common symptom of many other medical conditions that are not related to COVID-19.
Therefore, automatically classifying respiratory sounds for COVID-19 diagnostic is a non-trivial and challenging task.

During the pandemic, many crowdsourcing platforms (such as COUGHVID\footnote{https://coughvid.epfl.ch/}~\cite{orlandic2020coughvid}, COVID Voice Detector\footnote{https://cvd.lti.cmu.edu/}, and COVID-19 Sounds App\footnote{https://www.covid-19-sounds.org/en/})
have been designed to gather respiratory sound audios from both healthy and COVID-19 positive groups for the research purpose.
With these collected datasets, researchers in the artificial intelligence community have started to develop machine learning and deep learning based methods (\eg,~\cite{brown2020exploring,imran2020ai4covid,hassan2020covid,pahar2020covid,schuller2020detecting}) for cough classification to detect COVID-19.
Nevertheless, these methods share one common characteristic, that is they are all designed and trained in a fully-supervised way. 
On the one hand, the fully-supervised setting limits the applicability, effectiveness and impact of the collected datasets, since the method has to be trained and tested on the same dataset. This means additional datasets cannot be directly used to boost the predictive performance and the model is limited to the same source dataset.
On the other hand, such fully-supervised based classification methods inevitably need to rely on well-annotated cough sounds data.
The annotations are from either experts or user response surveys.
There are two inherent limitations of these annotation approaches:
    (i) \textit{Annotation Cost}: Annotation of a large-scale dataset comes at an expensive cost (both financial cost and human power cost). In addition, unlike the data labelling in other tasks such as image classification, the annotation of respiratory sounds requires specific knowledge from experts. This further aggravates the difficulty of obtaining accurate annotations.
    (ii) \textit{Privacy Concern}: Although directly asking participants to report their health status (\eg, whether the COVID-19 is tested positive or negative) during the respiratory sounds collection avoids annotation cost, the medical information is highly sensitive.
    Such privacy concerns also limit the distribution and publicity of gathered datasets. For example, some datasets have to be accessed by one-to-one legal agreements and specific licences.

\begin{figure}
    \centering
    \includegraphics[width=0.35\textwidth]{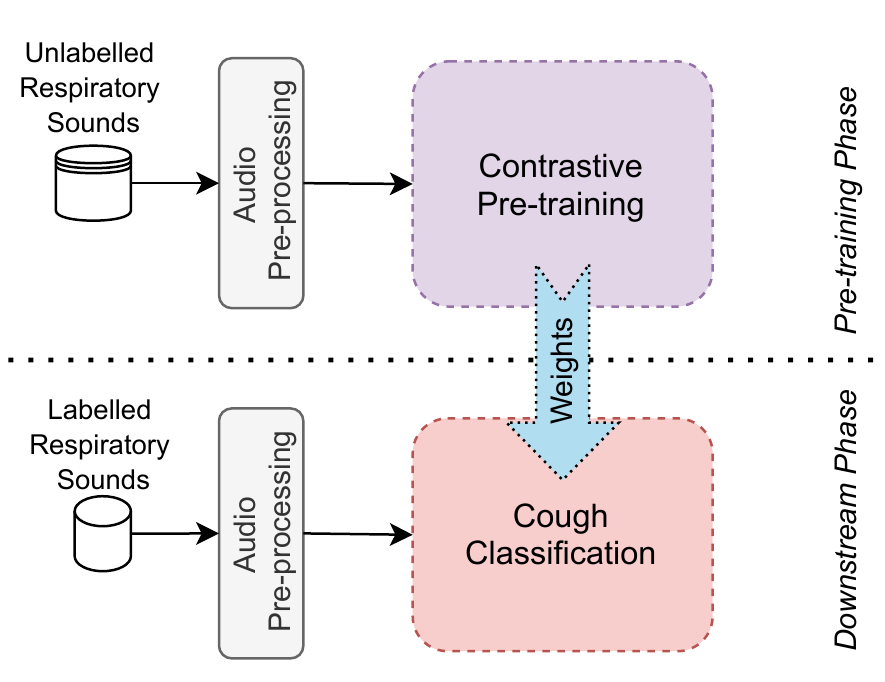}
    \caption{Concept illustration of the proposed framework. It consists of two phases: a contrastive pre-training phase (the upper part, described in Section~\ref{sec:p}) to learn representations from unlabelled respiratory data; and a downstream phase (the lower part, described in Section~\ref{sec:d}) for fine-tuning and performing cough classification for screening COVID-19.}
    \label{fig:overall_concept}
\end{figure}

In this work, to address the aforementioned shortcomings, we design a novel framework for COVID-19 cough classification, which can easily leverage large scale unlabelled respiratory sounds.
The concept of the proposed framework is illustrated in Figure~\ref{fig:overall_concept}.
Overall, it consists of two phases: a pre-training phase and a downstream phase.
The first phase is a self-supervised contrastive loss-based representation learning process with only unlabelled respiratory audios as training data.
The purpose is to train a feature encoder contrastively so that it can learn discriminative representations from large amount of unlabelled sounds.
In the downstream phase, the weights of the contrastive pre-trained feature encoder are transferred and fine-tuned on the labelled downstream dataset. 
Except for the loading of pre-trained weights, this phase is similar to other fully-supervised cough classification methods.
We pose the question of whether the contrastive pre-trained weights could help the downstream classification performance.
While the self-supervised contrastive representation learning has been successfully applied to other domains such as images~\cite{he2020momentum,chen2020simple}, speech~\cite{oord2018representation,jiang2020speech}, and general audios~\cite{saeed2020contrastive}, our work is the first attempt to explore self-supervised representations for respiratory sounds and COVID-19 cough classification.

For the audio feature encoder (pre-trained in the first phase and fine-tuned in the downstream phase), we adopt the popular Transformer architecture~\cite{vaswani2017attention} which has been proved effective in many other temporal data analysing tasks such as translation~\cite{vaswani2017attention,bert}, traffic prediction~\cite{xue2020trailer}, and event forecasting~\cite{wu2020hierarchically}.
Considering that the demographic distributions (\eg, age, gender, nationality of participants) of the pre-training data and the downstream dataset may be different, we explicitly design and introduce a random masking mechanism to improve the generalisation of the feature encoder. 
This mechanism randomly masks off some timestamps' signals in the input audio so that these masked values are removed from the attention calculation inside the Transformer. It could avoid the over-fitting on the pre-training dataset.
We also exploit applying the same random mechanism in the downstream phase in the experiments.
Furthermore, we also investigate different ensemble configurations with different feature encoder structure and random masking rates to further improve the classification performance.

In summary, our contributions are:
    (1)~We propose a novel framework based on contrastive pre-training to take advantage of unlabelled respiratory audios for representation learning. 
    To the best of our knowledge, this is the first paper using contrastive-based representation learning to leverage unlabelled data for COVID-19 cough classification.
    This framework provides a new perspective for the cough classification research.
    (2)~We design a random masking enabled Transformer structure as the feature encoder to learn the representations. Applying the random masking in the pre-training phase could provide effective and general representations, which further boosts the classification performance in the downstream phase.
    (3)~Through extensive experiments, we demonstrate that the proposed framework outperforms existing methods and we also discover the ensemble configuration that yields the best performance.

\section{Related Work}
\subsection{COVID-19 Cough Classification}

Machine learning as well as deep learning based methods have been introduced to automatically screen and diagnose various respiratory diseases~\cite{amoh2015deepcough,oletic2016energy,amoh2016deep,li2017design,bales2020can}.
As for the deep learning neural network architectures, there are two categories for COVID-19 cough classification.
The first category is Convolutional Neural Network (CNN) based.
Even though typical CNNs such as ResNet~\cite{he2016deep} and VGG~\cite{simonyan2014very} are originally proposed for processing images in computer vision, the pre-processing techniques (\eg, Mel Frequency Cepstral Coefficients (MFCC) and log-compressed mel-filterbanks) transform audio signals into 2D matrices and make it possible to directly apply CNN structures for audio analysis.
A CNN model with ResNet-18 as backbone is designed by Bagad~\etal~\cite{bagad2020cough}, whereas Schiller~\etal use an ensemble of CNNs to classify if a person is infected with COVID-19 in~\cite{schuller2020detecting}.
For respiratory sound classification, Brawn~\etal~\cite{brown2020exploring} combines hand-crafted features and VGGish~\cite{hershey2017cnn} (pre-trained on Audioset~\cite{gemmeke2017audio}) extracted deep learning features.

The second category is Recurrent Neural Network (RNN) based. Considering that audio data is inherently a type of temporal sequence data, modelling the recurrence dynamics~\cite{mouawad2021robust} is another technical road map for cough classification.
RNN and its variants Long Short Term Memory (LSTM) networks~\cite{hochreiter1997long}, Gated Recurrent Units (GRU)~\cite{chung2014empirical} are born for handling temporal sequence data.
Following this trend, Hassan~\etal~\cite{hassan2020covid} and Pahar~\etal~\cite{pahar2020covid} fully explore LSTM-based COVID-19 cough classification by researching and evaluating different sound features as input and LSTM hyperparameters.

The proposed framework in this work differs from the above summarised CNN or RNN based COVID-19 cough classification methods in the way of pre-training. 
Other existing methods with pre-training step depend on conventional fully-supervised pre-training so that large scale labelled data is required, whereas we introduce and design a contrastive self-supervised pre-training phase that only requires unlabelled data for pre-training.

\subsection{Self-Supervised Contrastive Learning}
The core idea of contrastive learning is to learn how to represent an input sample so that learned representations of positive pairs (samples considered to be similar) are much closer than representations of negative pairs (samples considered to be different) in the latent space. 
Recently, contrastive learning based self-supervised pre-training has been proved successful and effective to learn representations from unlabelled data in numerous work in other domains such as images~\cite{he2020momentum,chen2020simple} and speech~\cite{oord2018representation,jiang2020speech}.
Using such pre-trained representations could improve performance of downstream supervised tasks.

COLA~\cite{saeed2020contrastive}, proposed by Saeed~\etal, is the most relevant approach in the literature to our proposed framework. It is a contrastive learning framework to learn representations from general audios (i.e., Audioset) in a self-supervised manner.
However, there are two major differences between ours and COLA.
First, how to leverage unlabelled respiratory data for COVID-19 cough classification remains untouched in the literature.
We seek to develop a framework to learn representations for respiratory sounds based COVID-19 cough classification instead of representations for common audios in COLA. 
Second, COLA uses EfficientNet~\cite{tan2019efficientnet}, a CNN, as the feature encoder to extract representations from audio.
Instead, our model treats the audio as sequence data and utilises the popular Transformer~\cite{vaswani2017attention}, an effective architecture that has shown great promise in many other tasks, as the backbone for cough classification.
In addition, we propose a novel random masking mechanism to work together with the Transformer as the feature encoder in out framework.

\section{COVID-19 Respiratory Sound}\label{sec:3}

\subsection{Coswara Dataset}
Coswara dataset~\cite{sharma2020coswara} is part of Project Coswara\footnote{https://coswara.iisc.ac.in/} which aims to build a diagnostic tool for COVID-19 based on respiratory, cough, and speech sounds.
Upon until December 21\textsuperscript{st} 2020, there are 1,486 crowdsourced samples (collected from 1,123 males and 323 females) available at Coswara data repository\footnote{Coswara dataset can be accessed at https://github.com/iiscleap/Coswara-Data}.
The majority of the participants are from India (1,329 participants) and the rest participants are from other countries of five continents: Asia, Australia, Europe, North America, and South America.
Four types of sounds (breathing, coughing, counting, and sustained phonation of vowel sounds) are gathered from each participant.

\subsection{COVID-19 Sounds}

Similar to Coswara dataset, COVID-19 Sounds is another crowdsourcing based respiratory sound dataset. Audios are collected world-widely with a web-based app, an Android app, and an Apple app.
In our work, we choose the same curated dataset that is introduced and used in~\cite{brown2020exploring}.
After filtering out silent and noisy samples, in this released version of dataset\footnote{this dataset is available through one-to-one legal agreements.}, there are 141 COVID-19 positive audio recordings collected from 62 participants and 298 COVID-19 negative audio recordings from 220 participants.
Both coughs and breaths appear in these recordings.
Positive samples are from participants who claimed that they had tested positive for COVID-19.

\begin{figure*}
    \centering
    \includegraphics[width=0.75\textwidth]{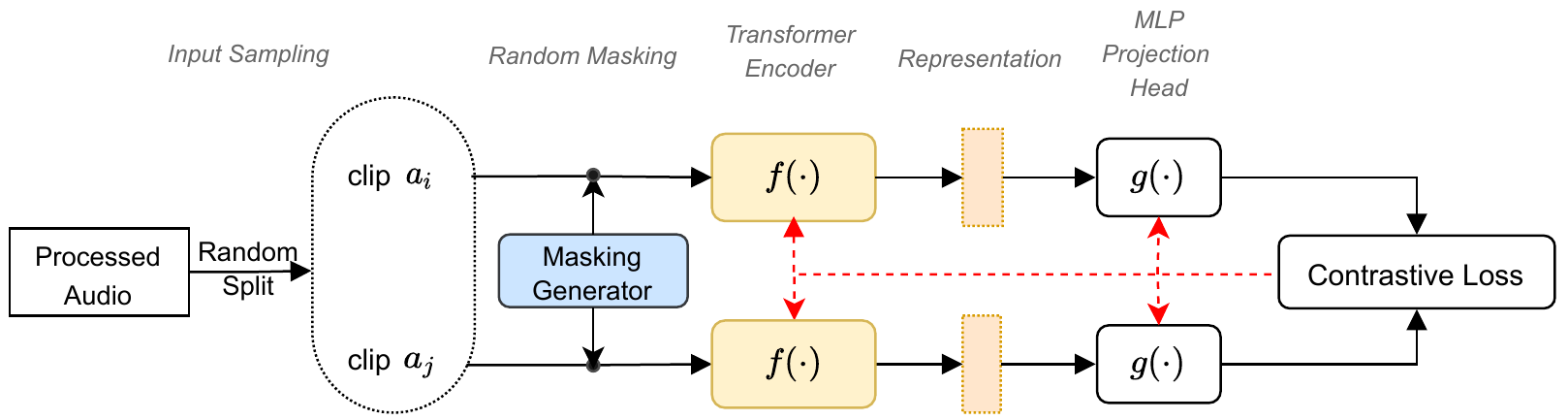}
    \caption{The pipeline of the contrastive pre-training of the proposed framework. The aim is to pre-train the feature encoder ($f(\cdot)$) which will be used in the downstream phase. A masking generator (the blue part) is used for random masking (detailed in Section~\ref{sec:mask}). Dash arrows indicate the back propagation operations in the pre-training.}
    \label{fig:contrastive_pretraining}
\end{figure*}

\section{Method}

As illustrated in Figure~\ref{fig:overall_concept}, the proposed method consists of two phases:
    (i) \textbf{Pre-training phase}: to pre-train the feature encoder with unlabelled audios through contrastive learning.
    (ii) \textbf{Downstream phase}: to fine-tune the trained feature encoder with an additional classifier for COVID-19 cough classification.
The details of these phases are given in the following subsections.

\subsection{Contrastive Pre-training}\label{sec:p}
The pipeline of the contrastive pre-training is given in Figure~\ref{fig:contrastive_pretraining}. The idea of contrastive learning can be summarised as: to encode audios into a latent space through the feature encoder so that the similarity of positive samples is larger than the negative samples in the latent space.
Therefore, three key components in this contrastive learning phase are: (1) how to obtain positive/negative samples; (2) how to design the feature encoder; and (3) how to measure the similarity in the latent space.

\subsubsection{Pre-processing and Sampling}\label{sec:pre}

The purpose of pre-processing is to read and transform each raw audio file into a matrix format which can be taken as input by the following feature encoder.
Mel Frequency Cepstral Coefficients (MFCC) and log-compressed mel-filterbanks have been widely used in the audio analysis~\cite{han2006efficient,chatrzarrin2011feature,pahar2020covid,hassan2020covid,brown2020exploring,saeed2020contrastive}.
Python Speech Features package\footnote{https://github.com/jameslyons/python\_speech\_features} is used for computing log-compressed mel-filterbanks in our framework.
After the pre-processing, each raw audio file is mapped to a feature $a \in \mathbb{R}^{N\times T}$, where $N$ stands for the number of frequency bins and $T$ indicates the total number of time frames in this audio. 

Since different audios in the dataset often have different lengths and different $T$ values after pre-processing, we apply a sliding window with window size $T_w$ to generate multiple clips for each processed audio. The sampling of positive and negative clips is then straightforward in our task. 
If clip $i$: $a_i\in \mathbb{R}^{N\times T_w}$ and clip $j$: $a_j\in \mathbb{R}^{N\times T_w}$ come from the same audio file, they are considered as a positive clip pair. On the contrary, if they are sampled from different audios, they form a negative pair.
It is worth noting that the sampling might be slightly different, depending on different pre-training datasets.
Let's say, for example, there are four respiratory sound files (fast/slow breathing sounds and deep/shallow cough sounds) gathered from the same participant. So, if two clips are from the same participant (any one or two from the four sound files), they are the positive pair. 
Overall, after contrastive learning, samples from the same person has a larger similarity in the latent space than samples from different persons.
Such positive/negative sampling does not involve any annotated labels regarding the health condition of participants.

\subsubsection{Feature Encoder with Random Masking}\label{sec:mask}

\begin{figure*}
    \centering
    \subfigure[Without Random Masking]{
    \includegraphics[width=.4\textwidth]{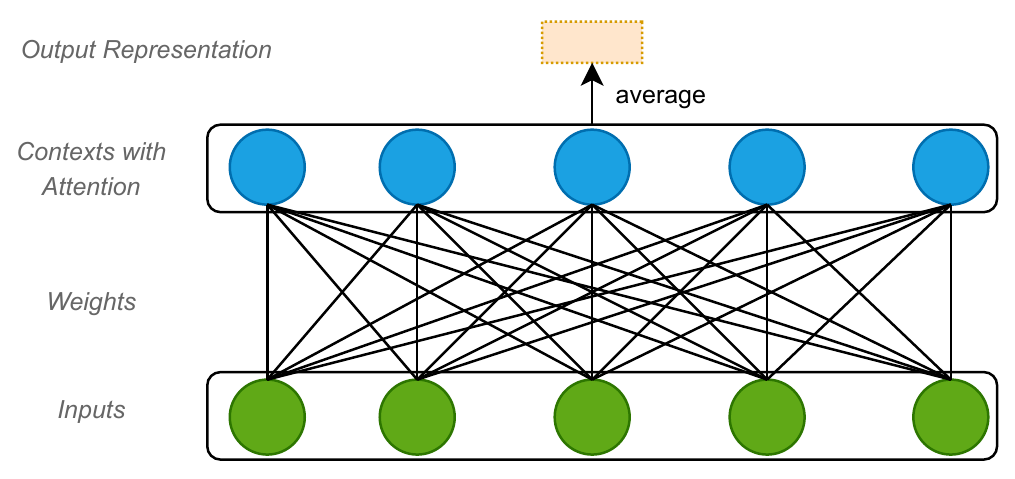}}
    \hspace{6ex}
    \subfigure[With Random Masking]{
    \includegraphics[width=.4\textwidth]{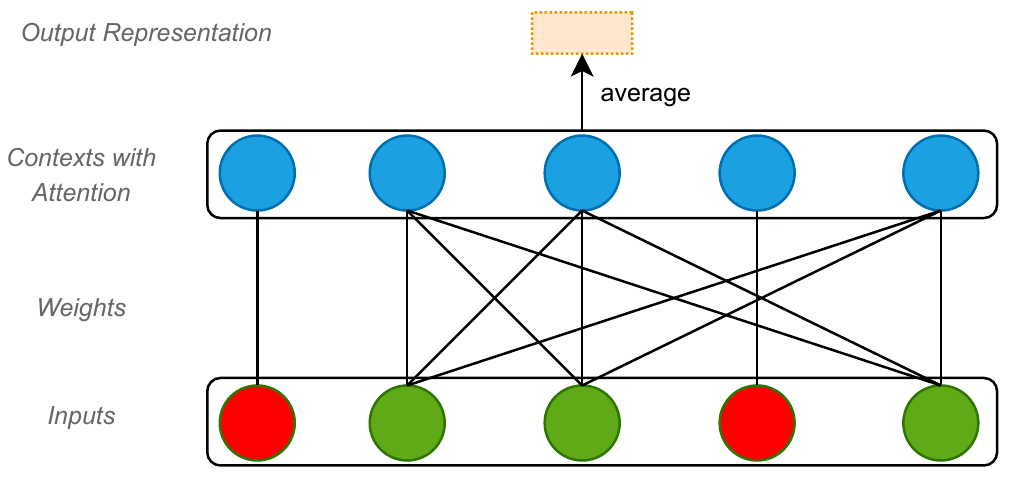}} 

    \caption{Illustration of random masking mechanism. (a) Without random masking; (b) With random masking. The masked inputs are shown as red circles and the masking rate is 40\% in this example.}
    \label{fig:masking}
\end{figure*}

The goal of the feature encoder is to embed each clip $a_i\in \mathbb{R}^{N\times T_w}$ into a representation vector $h_i\in \mathbb{R}^d$.
This step is formulated as:
\begin{equation}
    h_i = f (a_i; \mathbf{W}_f), \label{eq:1}
\end{equation}
where $f(\cdot)$ (light yellow box in Figure~\ref{fig:contrastive_pretraining}) represents the feature encoder and $\mathbf{W}_f$ is the trainable weights of the feature encoder.
Similarly, $h_j \in \mathbb{R}^d$ for clip $a_j$ is obtained.
The dimension of a representation vector is $d$.
In the proposed framework, we select the popular and effective Transformer structure~\cite{vaswani2017attention} as the feature encoder $f(\cdot)$.

As shown in Figure~\ref{fig:masking}(a), the typical Transformer structure models the input sequence ($a_i$ is considered as a $T_w$ time steps sequence and each time step is $N$ dimension) through the attention mechanism.
For each time step, the scaled dot-product attentions~\cite{vaswani2017attention} for every other time step are calculated.
Given that we need to transfer the pre-trained weights of $f(\cdot)$ to the downstream dataset, such a densely calculated attention mechanism might cause over-fitting on the pre-training dataset. 
To this end, we introduce a random masking mechanism (see Figure~\ref{fig:masking}(b)) to make the feature encoder robust.
For a respiratory sound, the feature at each time step might not be always meaningful. A collected sound sample often contains noises such as a short pause between two coughs. This also motivates us to design this random masking.

The masking generator (the blue box in Figure~\ref{fig:contrastive_pretraining}) generates a masking matrix $M$ with a specific masking rate (this rate is adjustable hyperparameter). 
Based on the masking matrix and the masking rate, some of the inputs are randomly masked and removed from the attention calculation in the Transformer. For example, in Figure~\ref{fig:masking}(b), with a 40\% masking rate, 2 time steps (shown in red circles) out of 5 time steps are masked.
With the random masking, Equation~\eqref{eq:1} is involved to:
\begin{equation}
    h_i = f(a_i, M_i; \mathbf{W}_f),
\end{equation}
where $M_i$ is the masking matrix for clip $a_i$.

\subsubsection{Contrastive Learning}

As suggested in many other contrastive learning methods such as~\cite{chen2020simple,saeed2020contrastive}, a projection head $g(\cdot)$ (see Figure~\ref{fig:contrastive_pretraining}) is applied to map representations (\eg, $h_i$ and $h_j$) to the latent space where the similarity is measured.
To measure the similarity, two metrics are used in the literature:
\begin{itemize}
    \item Cosine Similarity: this metric is commonly used visual representation learning such as~\cite{chen2020simple,he2020momentum}.
    The similarity of a clip pair $\text{sim}(a_i, a_j)$ is calculated by:
    \begin{equation}
        \text{sim}(a_i, a_j) = \frac{g(h_i)^\top\cdot g(h_j)}{\left \| g(h_i)) \right \|\left \| g(h_j)) \right \|}
    \end{equation}
    \item Bilinear Similarity: this similarity has been used in~\cite{oord2018representation,saeed2020contrastive}. The similarity of a clip pair is given as:
    \begin{equation}
        \text{sim}(a_i, a_j) = g(h_i)^\top \mathbf{W}_s g(h_j),
    \end{equation}
    where $ \mathbf{W}_s$ is the bilinear parameter.
\end{itemize}
Specifically, we conduct an experiment to compare the performance of these two types of similarity metrics in Section 5.3.

The loss function used in this phase for contrastive learning is a multi-class cross-entropy function working together with the similarity metric.
During the training in this phase, each training instance (consists of two clips from the same participant) in a batch is a positive pair. 
Clips from different training instances (from different participants) formulate negative pairs. 
Each training instance is then considered as a unique "class" (a unique participant), so the multi-class cross-entropy is applied.
This loss function is calculated over the batch (batch size $B$ and $2B$ indicates the total number of clips in a batch as we have two clips for each training instance) and modelled as:
\begin{equation}
    \mathcal{L}_{\text{contrastive}} = -\log \frac{\exp (\text{sim}(a_i, a_j)/\tau) }{\sum_{k=1}^{2B}\exp(\text{sim}(a_i, a_k)/\tau)}, k\neq i, \label{eq:c_loss}
\end{equation}
where $\tau$ denotes the temperature parameter for scaling.
Note that in Equation~\eqref{eq:c_loss}, $a_i$ and $a_j$ is the positive pair whereas $a_i$ and $a_k$ ($k \neq i$) are all the negative pairs.

\subsection{Downstream Cough Classification}\label{sec:d}

\begin{figure}
    \centering
    \includegraphics[width=.45\textwidth]{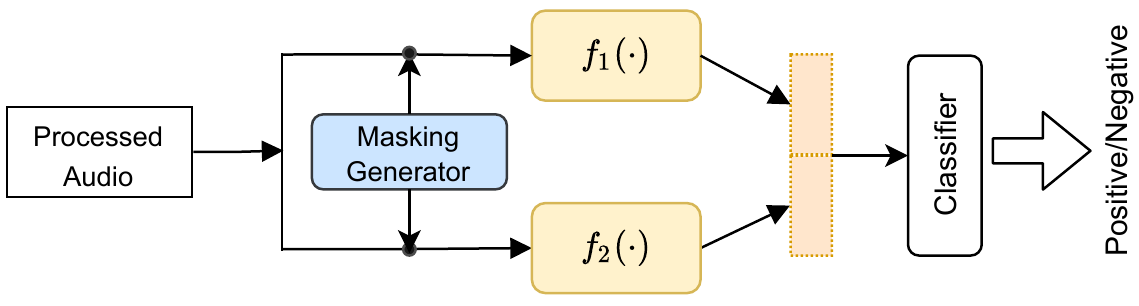}
    \caption{Illustration of an ensemble network structure for cough classification in the downstream phase. The masking generator is also included for random masking.}
    \label{fig:downstream}
\end{figure}

In the downstream phase, a straightforward network architecture is the feature encoder ($f(\cdot)$) with an additional classifier. 
The feature encoder is initialised with pre-trained weights ($\mathbf{W}_f$) in the previous phase and takes pre-processed audio clip as input. The encoded feature $h_i$ is then passed to the classifier.
The classifier is a fully-connected layer with $d$ (feature dimension of the encoded feature $h_i$) input neurons and one output node with sigmoid activation function to output the probability and indicate whether the input respiratory sound clip is COVID-19 positive (probability larger than a threshold, \eg\ 0.5) or negative (probability smaller than the threshold).
The network is fine-tuned with labelled data end-to-end with typical binary cross-entropy classification loss.

Based on this straightforward architecture, we also explore and design an advanced architecture (illustrated in Figure~\ref{fig:downstream}) with the random masking mechanism and an ensemble structure.
The motivation of introducing an ensemble structure for classification is related to the random masking.
Since the masking matrix is generated randomly, the two branches in the ensemble structure ($f_1(\cdot)$ and $f_2(\cdot)$ in Figure~\ref{fig:downstream}) would have different masked time steps, which leads the feature encoders to model the input audio and yield encoded features from different perspectives.
Thus, the ensemble structure and the random masking mechanism is a harmonised match and beneficial to each other.
Unlike the above straightforward architecture, the classifier in the ensemble architecture has $2d$ input neurons as it takes the concatenated feature (the concatenation of two encoded vectors from $f_1(\cdot)$ and $f_2(\cdot)$) as input. 

Note that both $f_1(\cdot)$ and $f_2(\cdot)$ are initialised with the same pre-trained weights from the contrastive pre-training phase.
During the fine-tuning process, both feature encoders may be updated differently.

\section{Evaluation}
In this work,  we focus on investigating the following research questions:
    (1)~RQ1: Whether introducing the contrastive pre-training has better performance than conventional fully-supervised setting
    and which similarity metric has the better performance in our cough classification task?
    (2)~RQ2: Does the random masking mechanism help the cough classification performance and what is the most suitable masking configuration?
    (3)~RQ3: By introducing the ensemble framework in the downstream phase, could we achieve a further improvement regarding the cough classification performance?

\subsection{Experimental setup}

\subsubsection{Data Processing}
As introduced in Section~\ref{sec:3}, we focus on two public COVID-19 respiratory datasets. 
Considering that Coswara dataset~\cite{sharma2020coswara} has more participants and contains more audio samples than COVID-19 Sounds dataset~\cite{brown2020exploring}, the Coswara dataset is adopted as the pre-training dataset in this work.
Note that for this pre-training dataset, the annotated labels (indicating whether the user is COVID-19 positive or negative) are not used.
Furthermore, since this work is more about respiratory sounds, breathing sounds and cough sounds are selected for pre-processing and sampling (detailed in Section~\ref{sec:pre}), whereas audios of sustained phonation of vowel sounds and counting sounds are ignored in the pre-training phase.
Consequently, COVID-19 Sounds is used as the dataset in the downstream phase.
To be more specific, in the downstream phase, the whole COVID-19 Sounds dataset is randomly divided into the training set (70\%), validation set (10\%), and testing set (20\%).
For each raw audio sample, the same pre-processing procedure (described in Section~\ref{sec:pre}) is applied as well.

\begin{figure*}
    \centering
    \subfigure[ROC-AUC]{
    \includegraphics[width=.24\textwidth]{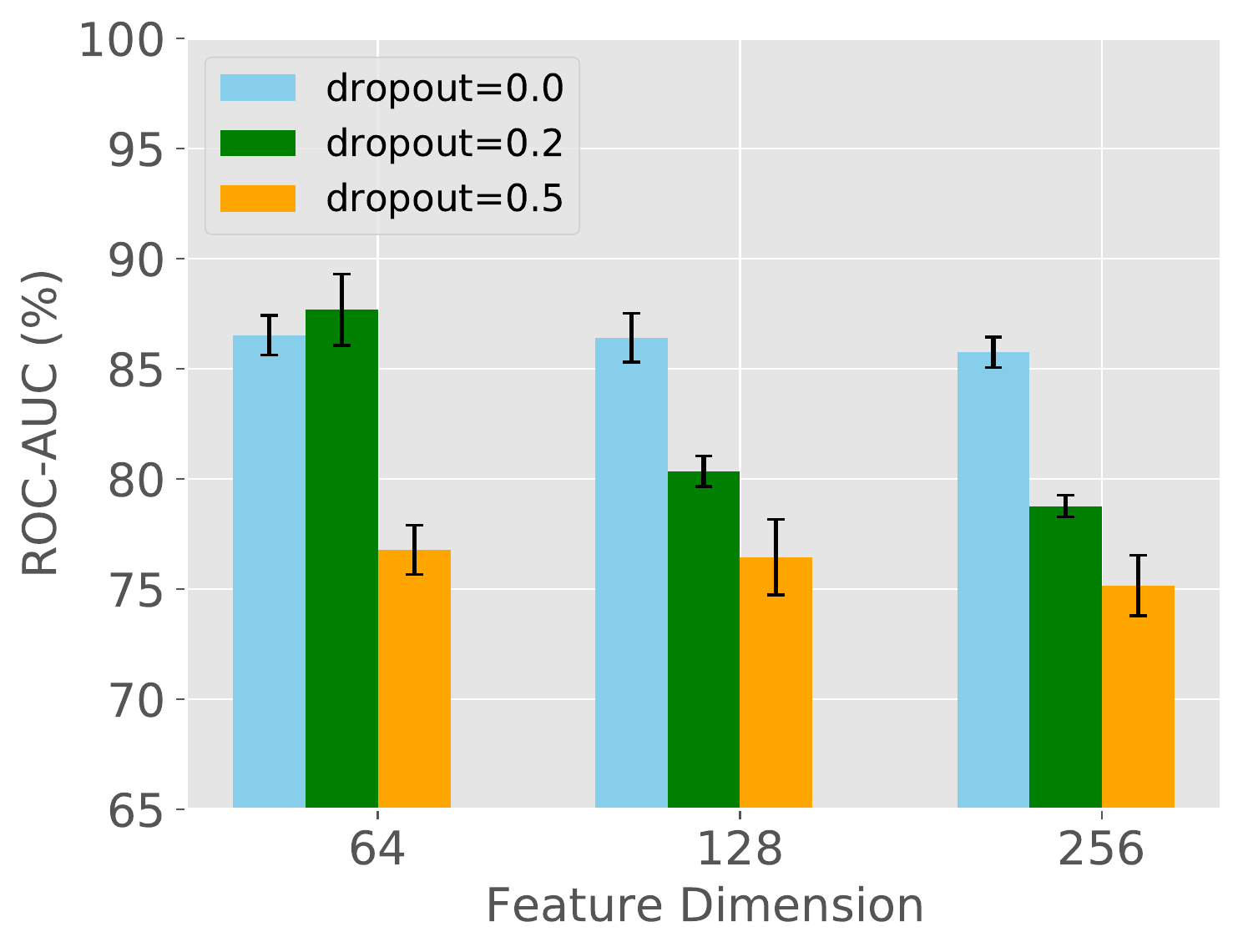}} 
    \subfigure[Recall]{
    \includegraphics[width=.24\textwidth]{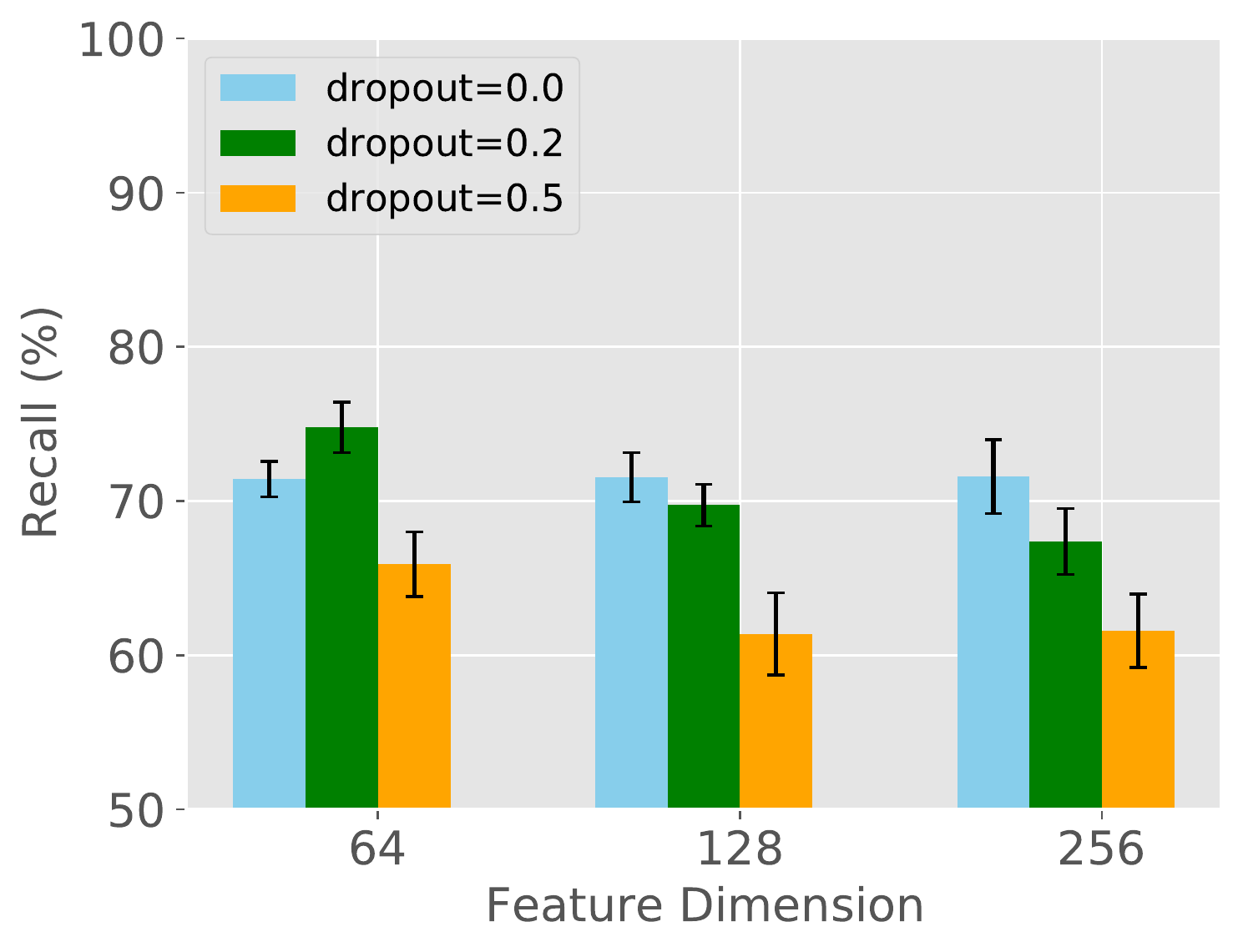}} 
    \subfigure[Precision]{
    \includegraphics[width=.24\textwidth]{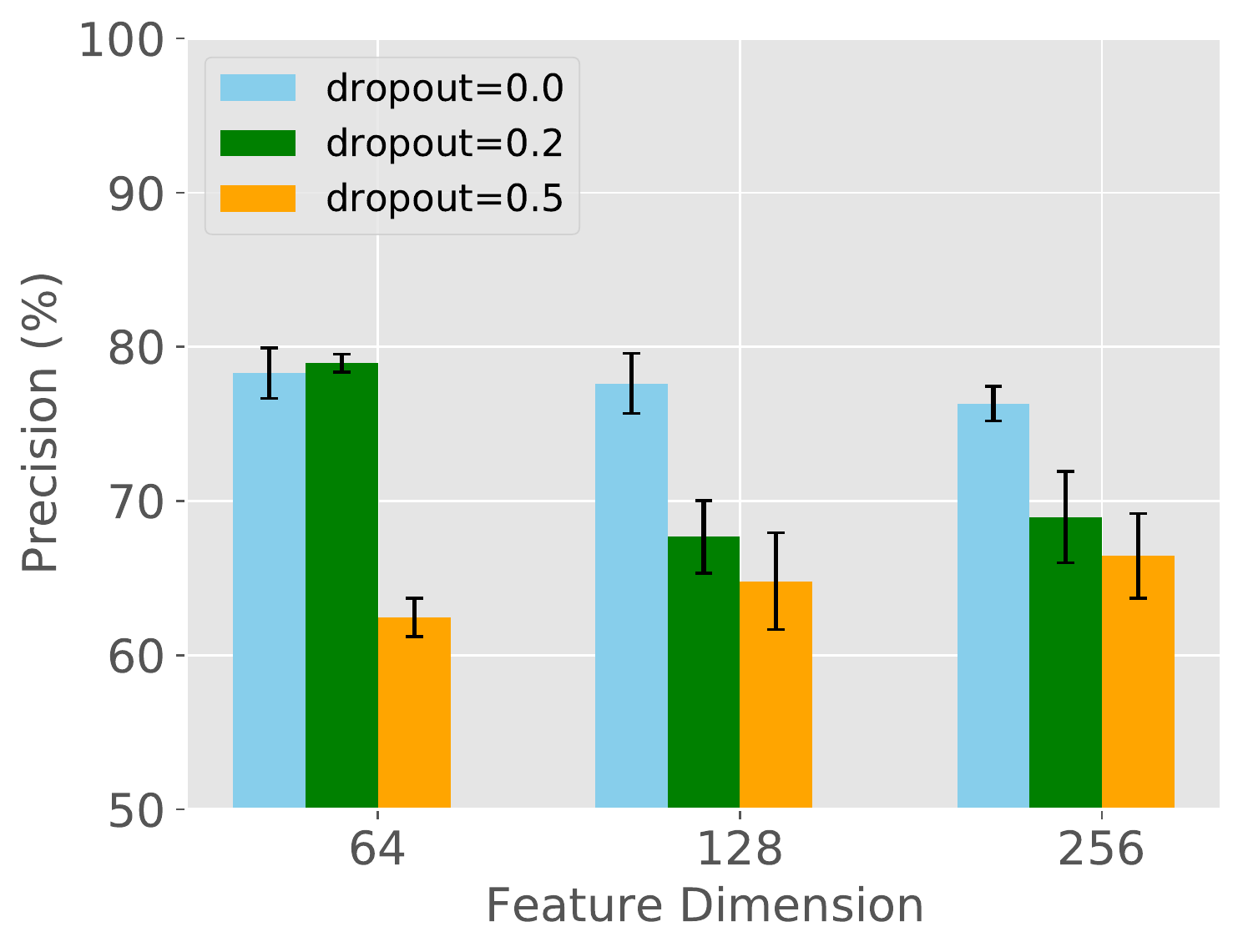}} 
    \subfigure[Accuracy]{
    \includegraphics[width=.24\textwidth]{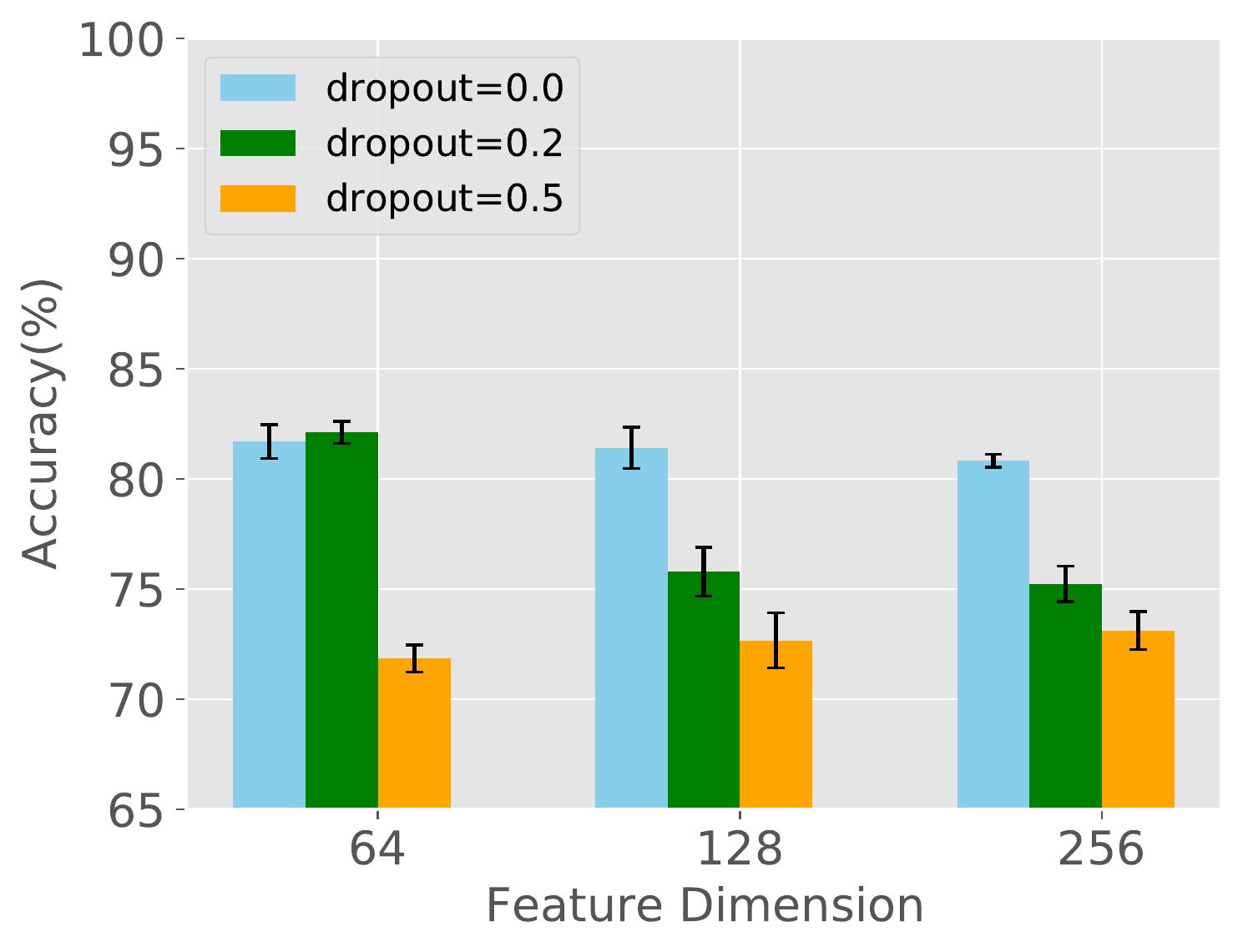}}
    \caption{The performance of different hyperparameter settings on the validation set.}
    \label{fig:hyper}
\end{figure*}

\begin{table*}[]
\centering
\caption{Results (on the testing set) of different models and configurations. For each result, the standard deviation is reported in a bracket.}
\begin{tabular}{|l|c|c|c|c|c|c|c|c|}
\hline
Model      & Self-supervise & Pre-train      & Fine-tune & ROC-AUC      & Recall       & Precision    & Accuracy     & Average F1 \\ \hline\hline
            & $\times$ & $\times$         & N/A         & 76.14 (0.21) & 53.19 (2.29) & 73.17 (1.87) & 74.88 (0.22) & 61.60      \\
VGGish      & $\times$ & \checkmark & $\times$          & 85.02 (1.73) & 67.42 (2.06) & 78.55 (2.21) & 80.71 (1.32) & 72.56      \\
            & $\times$ & \checkmark              & \checkmark         & 87.34 (1.14) & 69.49 (1.44) & \textbf{83.15 (1.46)} & 83.12 (0.31) & 75.71      \\ \hline
GRU         & $\times$ & $\times$          & N/A         & 84.43 (0.88) & 65.60 (1.43) & 82.67 (1.89) & 81.76 (0.39) & 73.15      \\
Transformer & $\times$ & $\times$          & N/A         & 87.60 (0.71) & 71.53 (1.12) & 80.64 (1.19) & 82.73 (0.41) & 75.81      \\ \hline
\multirow{2}{*}{GRU-CP}         & \checkmark & \checkmark  & $\times$ & 83.20 (0.43) & 63.43 (1.82) & 78.63 (1.22) & 79.63 (0.31) & 70.22 \\
            & \checkmark & \checkmark  & \checkmark         & 87.08 (0.35) & 71.72 (2.53) & 81.61 (2.26) & 83.15 (0.32) & 76.35      \\ \hline
\multirow{2}{*}{Transformer-CP} & \checkmark & \checkmark  & $\times$ & 84.34 (0.71) & 64.94 (1.80) & 78.56 (1.40) & 80.02 (0.42) & 71.10 \\
            & \checkmark & \checkmark  & \checkmark         & \textbf{88.83 (0.53)} & \textbf{73.07 (0.65)} & {81.99 (0.92)} & \textbf{83.74 (0.39)} & \textbf{77.27}     \\ \hline
\end{tabular}
\label{tab:cp_results}
\end{table*}

\subsubsection{Implementation Details}

In the pre-processing, the shape of a processed clip $a_i$ is $\mathbb{R}^{64\times96}$ as the number of mel-spaced frequency bins $N$ is set to 64 and the sliding window size $T_w$ is 96 which corresponds to 960 ms.
The feature dimension $d$ is set to 64.
In the contrastive pre-training phase, the batch size $B$ is selected as a large number (1024). 
As suggested by other contrastive learning methods (\eg,~\cite{chen2020simple}), the contrastive learning benefits from larger batch sizes (within GPU capacity) as a larger batch allows the model to compare the positive pair against more negative pairs.
In the downstream network, the dropout is also applied to avoid over-fitting in the end-to-end fine-tuning process.
The validation set in the downstream dataset is used for tuning hyperparameter $d$ (the feature dimension of the feature encoder) and the dropout rate.
The batch size is 128 for the downstream phase. 
All experiments (both the contrastive pre-training and the downstream phases) are trained with Adam optimiser~\cite{kingma2014adam} (a 0.001 initial learning rate with \textit{ReduceLROnPlateau}\footnote{https://pytorch.org/docs/stable/optim.html\#torch.optim.lr\_scheduler.ReduceLROnPlateau} decay setting) and executed on a desktop with an NVIDIA GeForce RTX-2080 Ti GPU with  PyTorch.

\subsubsection{Evaluation Metrics}
To evaluate the performance of different methods, several standard classification evaluation metrics including the Receiver Operating Characteristic - Area Under Curve (ROC-AUC), Precision, and Recall are selected. 
In our experiments, we report the average performance as well as the standard deviation of 5 runnings of each method or configuration.
In addition, we report the average F1 score which is calculated based on the average Precision and average Recall.

\subsection{Hyperparameter Fine-Tuning}

To investigate how the dimension of encoded feature and dropout rate  influence the classification performance, all the combination of the following hyperparameter values (within reasonable ranges) are evaluated on the validation set:
    (1) Feature Dimension: $[64, 128, 256]$;
    (2) Dropout Rate: $[0.0, 0.2, 0.5]$
(resulting in 9 combinations in total). 
Please note that we also run 5 times for each combination.
Figure~\ref{fig:hyper} shows the average performance of four metrics and error bars indicate the standard deviations.
Based on these validation results, $d=64$ and a 0.2 dropout rate achieve the best validation performance 
and are used for the rest experiments. 
Since the feature encoder structure should be identical in both the pre-training and the downstream phases, the same hyperparameter setting is also applied in the pre-training phase.

\begin{table*}[]
\centering
\caption{Results (on the testing set) of two types of similarity metrics that are used in the contrastive pre-training phase.}
\begin{tabular}{|l|c|c|c|c|c|c|} \hline
Model & Similarity Metric & ROC-AUC      & Recall       & Precision    & Accuracy     & Average F1 \\ \hline\hline
\multirow{2}{*}{GRU-CP}         & Cosine & 87.08 (0.35) & 71.72 (2.53) & 81.61 (2.26) & 83.15 (0.32) & 76.35 \\
      & Bilinear        & 87.25 (0.65) & 71.99(1.97)  & 82.62 (1.92) & 83.65 (0.32) & 76.94      \\ \hline
\multirow{2}{*}{Transformer-CP} & Cosine & 87.02 (0.53) & 67.79 (2.12) & 81.73 (1.54) & 82.06 (0.33) & 74.11 \\
      & Bilinear        & 88.83 (0.53) & 73.07 (0.65) & 81.99 (0.92) & 83.74 (0.39) & 77.27      \\ \hline
\end{tabular}
\label{tab:similarity_loss}
\end{table*}

\begin{table*}[]
\centering
\caption{Cough classification results (on the testing set) of different masking rates used in the contrastive pre-training phase.}
\begin{tabular}{|l|c|c|c|c|c|c|} \hline
Model          & Masking Rate (CP) & ROC-AUC      & Recall       & Precision    & Accuracy     & Average F1 \\ \hline\hline
Transformer    & NA           & 87.60 (0.71) & 71.53 (1.12) & 80.64 (1.19) & 82.73 (0.41) & 75.81      \\\hline
               & 0\%          & 88.83 (0.53) & 73.07 (0.65) & 81.99 (0.92) & 83.74 (0.39) & 77.27      \\
               & 25\%         & 89.17 (0.13) & 73.05 (0.83) & 82.29 (1.13) & 83.84 (0.38) & 77.40       \\
Transformer-CP & 50\%         & \textbf{89.42 (0.51)} & \textbf{73.09 (0.43)} & \textbf{83.26 (0.28)} & \textbf{84.26 (0.12)} & \textbf{77.84}      \\
               & 75\%         & 89.13 (0.95) & 72.66 (1.20) & 82.01 (1.52) & 83.62 (0.41) & 77.05      \\
               & 100\%        & 88.37 (0.57) & 72.41 (0.64) & 81.65 (0.82) & 83.41 (0.41) & 76.75     \\ \hline
\end{tabular}
\label{tab:cp_masking}
\end{table*}

\subsection{Contrastive Pre-training Performance}

\subsubsection{Methods for Comparison}
To evaluate the performance of contrastive pre-training and the Transformer feature encoder, we compare \textbf{Transformer-CP} (the suffix \textit{-CP} means the method is contrastive pre-training enabled) with several methods with multiple configurations. 
Other methods being compared include VGGish/GRU/Transformer (without contrastive pre-training) and GRU-CP.
Recurrent Neural Networks (RNNs) are designed for handling sequence data and have been adopted for COVID-19 cough classification research in~\cite{pahar2020covid,hassan2020covid}. So, the GRU~\cite{chung2014empirical} is also included in the comparison.
VGGish~\cite{hershey2017cnn} is a popular convolutional neural network for audio classification. A pre-trained version\footnote{Pre-trained weights of VGGish can be found at https://github.com/tcvrick/audioset-vggish-tensorflow-to-pytorch/releases.} that is pre-trained on a large scale general audio dataset Audioset~\cite{gemmeke2017audio} is also widely used in the community.
Such a pre-trained VGGish has also been applied in~\cite{brown2020exploring} to extract features for COVID-19 cough classification.
Note that the pre-training of VGGish is a conventional fully-supervised pre-training with labelled data, which is different from our contrastive pre-training. 
The configuration of with or without the pre-training is summarised in the third column of Table~\ref{tab:cp_results}. In addition, the second column indicates the pre-training setting. A \checkmark\ means that the proposed self-supervised contrastive pre-training is applied.
For example, both the second and third columns are \checkmark\ for our Transformer-CP.

\subsubsection{Performance Comparison}
The experimental results of the above methods are reported in Table~\ref{tab:cp_results}. 
To be more specific, for methods using pre-trained weights (either contrastive pre-training or conventional pre-training for VGGish), we also explore the fine-tuning option. In the fourth column of Table~\ref{tab:cp_results}, a $\times$ represents the pre-trained weights $\mathbf{W}_f$ are frozen and not be updated in the downstream phase, whereas a \checkmark indicates $\mathbf{W}_f$ is allowed to be updated.

According to the table, the proposed Transformer-CP with fine-tuning achieves the best performance (shown in bold) against all the other methods. There are several additional findings that can be noticed from the table.
First, without pre-training, VGGish has the worst performance (the first row) compared to GRU and Transformer. 
Using pre-trained VGGish weights (without fine-tuning) provides almost 6\% accuracy gain, which indicates that the pre-trained VGGish representation is well-trained and powerful.
For all configurations that using frozen pre-trained representations (the second, sixth, and eighth rows), although VGGish (the second row) is the top performer, the performance of our Transformer-CP (the eighth row) is very close to VGGish.
This is remarkable as it shows that our contrastive pre-trained (on a smaller scale and unlabelled dataset) self-supervised feature representation is competitive with a well-trained fully-supervised VGGish representation (pre-trained on a much larger scale and well-annotated Audioset).
Second, the fine-tuning in the downstream task is important for all pre-trained models, which is as expected. For both the conventional pre-trained VGGish and contrastive pre-trained GRU/Transformer, the fine-tuning could improve the accuracy by around 3\%.
Third, if we compare GRU vs. Transformer and GRU-CP vs. Transformer-CP, the Transformer-based methods outperform GRU-based methods consistently. This justifies the selection of Transformer as the feature encoder in the proposed framework.
Overall, the results show that the proposed framework with contrastive pre-training achieves a superior performance of cough classification.

\subsubsection{Different Similarity Metrics}
In Table~\ref{tab:similarity_loss}, two similarity metrics in contrastive learning are compared. 
For a fair comparison, two different feature encoder structures, GRU-CP and Transformer-CP, are explored.
As shown in the table, using bilinear similarity achieves consistent better performance with both structures on all evaluation metrics, which demonstrates that the bilinear similarity is more suitable for our cough classification task. 

\subsection{Random Masking Performance}

In this part of the experiments, we turn to research on the proposed random masking mechanism and different masking rates in the contrastive pre-training phase.
The experiment guideline for this part is: we pre-train several Transformer-CPs with multiple masking rates (0\% to 100\%) and then the pre-trained models are fine-tuned in the downstream phase. 
The cough classification performance of these models are listed in Table~\ref{tab:cp_masking}.
Please note that in the downstream phase, we do not apply the ensemble architecture so that there is no random masking in the downstream phase for results reported in the table. 

As a baseline for comparison, we also include the performance of Transformer (without any pre-training) in Table~\ref{tab:cp_masking}. In general, all pre-trained models yield better results than the baseline Transformer and 50\% masking outperforms other masking rates.
When the masking rate is increasing from 0\% (no masking at all) to 50\%, we can witness a performance gain from the table.
However, when the masking rate is too large (\eg, 75\% and 100\%), the performance decreases.
This is not surprising. For example, in the extreme 100\% masking case, all the inputs are masked, which means there is no attention between any time steps. As a result, the 100\% masking has the worst performance among different masking rate settings.

\begin{table*}[]
\centering
\caption{Cough classification results (on the testing set) of different ensemble configurations.}
\begin{tabular}{|cc|c|c|c|c|c|} \hline
Ensemble\_1 & Ensemble\_2 & ROC-AUC & Recall & Precision & Accuracy & Average F1 \\ \hline\hline
VGGish & Transformer & 87.58 (0.73) & 70.30 (1.05) & 82.46 (1.44) & 83.10 (0.26) & 75.90 \\
GRU & Transformer & 87.24 (0.59) & 72.56 (1.81) & 81.10 (1.54) & 83.20 (0.38) & 76.59 \\
GRU-CP & Transformer-CP & 88.90 (0.38) & 72.77 (1.85) & 83.59 (1.62) & 84.04 (0.35) & 77.81 \\ \hline
\end{tabular}
\label{tab:all_ens}
\end{table*}

\begin{table*}[]
\centering
\caption{Results (on the testing set) of combining different masking rates with ensembles in the downstream phase.}
\begin{tabular}{|cc|c|c|c|c|c|c|} \hline
Ensemble\_1 & Ensemble\_2 & Masking (DS) & ROC-AUC & Recall & Precision & Accuracy & Average F1 \\ \hline\hline
 &  & 0\% & 88.77 (0.87) & 71.98 (0.67) & 82.81 (0.59) & 83.75 (0.38) & 77.02 \\
 &  & 25\% & 89.02 (0.55) & 71.93 (1.01) & 82.85 (0.48) & 84.15 (0.20) & 77.00 \\
Transformer-CP & Transformer-CP & 50\% & \textbf{90.03 (0.41)} & \textbf{73.24 (0.22)} & \textbf{84.57 (1.24)} & \textbf{84.43 (0.25)} & \textbf{78.50} \\
 &  & 75\% & 89.22 (0.24) & 72.12 (0.32) & 83.17 (0.21) & 84.03 (0.58) & 77.25 \\
 &  & 100\% & 89.55 (0.91) & 71.21 (1.79) & 82.85 (2.33) & 83.93 (0.17) & 76.59 \\ \hline
\end{tabular}
\label{tab:t_ens}
\end{table*}

\begin{table*}[]
\centering
\caption{Comparison of inference speed of different model and configurations. Each method is benchmarked on the same NVIDIA GeForce RTX-2080 Ti GPU.}
\begin{tabular}{|c|c|ccc|c|} \hline
Ensemble & Model & Self-supervise & Pre-train & Masking Rate (DS) & Inference time ($10^{-6}$ seconds) \\ \hline
\multirow{6}{*}{$\times$} & \multirow{2}{*}{VGGish} & $\times$ & $\times$ & N/A & 6.39 \\
 &  & $\times$ & \checkmark  & N/A & 6.39 \\ \cline{2-6}
 & GRU & $\times$ & $\times$ & N/A & 8.34 \\
 & Trasformer & $\times$ & $\times$ & N/A & 8.48 \\ \cline{2-6}
 & GRU-CP & \checkmark  & \checkmark  & N/A & 8.60 \\
 & Transformer-CP & \checkmark  & \checkmark  & N/A & 8.52 \\ \hline\hline
\multirow{8}{*}{\checkmark } & VGGish + Transformer & $\times$ & $\times$ & N/A & 12.58 \\
 & GRU + Transformer & $\times$ & $\times$ & N/A & 14.26 \\
 & GRU-CP + Transformer-CP & \checkmark  & \checkmark  & N/A & 14.36 \\ \cline{2-6}
 & \multirow{5}{*}{Transformer-CP + Transformer-CP} & \checkmark  & \checkmark  & 0\% & 18.86 \\
 &  & \checkmark  & \checkmark  & 25\% & 24.53 \\
 &  & \checkmark  & \checkmark  & 50\% & 27.56 \\
 &  & \checkmark  & \checkmark  & 75\% & 32.36 \\
 &  & \checkmark  & \checkmark  & 100\% & 18.64 \\ \hline
\end{tabular}
\label{tab:speed}
\end{table*}

\subsection{Ensembles Performance}

In this section, we focus on exploring different ensembles.
Table~\ref{tab:all_ens} summarises three ensemble methods. The first two are the ensemble of our base Transformer feature encoder and other feature encoder structures (VGGish and GRU). No pre-trained weights are applied to these two ensembles.
The third ensemble combines GRU-CP and Transformer-CP with contrastive pre-trained weights.
By jointly comparing results given in Table~\ref{tab:cp_results} and Table~\ref{tab:all_ens}, it can be seen that the ensemble version demonstrates a better ability than a single feature encoder based method.

Moreover, we investigate networks where the random masking is incorporated with the ensemble architecture (as shown in Figure~\ref{fig:downstream}).
For the ensembles presented in Table~\ref{tab:t_ens}, both branches are set as Transformer-CP. 
We manipulate the masking rate in the downstream phase (rates given in the \textit{Masking (DS)} column).
In addition, the pre-trained weights of the top performer in Table~\ref{tab:cp_masking} (with 50\% contrastive pre-training masking rate) are used for these ensembles. 
Similar to the masking in the contrastive pre-training phase, 50\% masking rate in the downstream phase also performs better than other masking rates.
The above results confirm that the proposed ensemble architecture with the random masking could further improve the classification performance. 

\subsection{Inference Speed}


Table~\ref{tab:speed} lists the inference time (for one input instance) of each model or configuration. 
Since the fine-tuning does not affect the inference time, the fine-tuning configuration is removed for comparison in the table.
Generally, for three different base feature encoder structures, the inference time of Transformer is on par with GRU, whereas VGGish leads Transformer/GRU by a small margin (around 0.002 milliseconds only). 
Although Transformer includes attention computation, it processes each time step in the input sequence in parallel, whereas GRU has to process each time step recurrently.
This might explain the similar computation cost between Transformer and GRU.
From the table, we notice also that using contrastive pre-trained weights does not introduce a longer time for inference.
This is as expected as the major difference between Transformer and Transformer-CP (or GRU vs. GRU-CP) is whether loading the pre-trained weights. This weights initialisation process almost has no influence on the inference speed.

An interesting and surprising finding is about the inference time of using different downstream random masking rates (the last five rows of Table~\ref{tab:speed}). 
In theory, a larger masking rate should run faster as more time steps are masked and not used in attention calculation.
According to the table, however, 75\% rate has the largest inference time and 0\% and 100\% are all smaller than the rest masking rates.
This can be explained by the implementation of the masking generator. 
In the implementation, the default masking matrix is an all-ones matrix or an all-zeros matrix (only used for masking rate 100\%), where 0 means being masked and vice versa. 
For a given masking rate, 1 will be updated to 0 in the matrix through a for loop.
This loop operation takes longer if more elements need to be updated (\eg, 75\% rate), which causes the larger inference time for the 75\% setting.
Overall, even the largest time cost in the table is only ${32.36}\times10^{-6}$ seconds (around 0.03 milliseconds). 
Such a low time cost would not be a bottleneck or limit the application of the proposed framework.

From another point of view, without the proposed contrastive pre-training, multiple models need to be trained if multiple datasets are available.
As a result, training time will be done per model without domain transfer, which is a potential bottleneck for large-scale deployments.
However, our proposed framework is able to address this training bottleneck through the contrastive pre-training phase.

\section{Conclusion}
In this paper, we propose a novel framework for respiratory sounds based COVID-19 cough classification. This study appears to be the first study to leverage unlabelled respiratory audios in the area.
In order to do so, we introduce a contrastive pre-training phase in which the Transformer-based feature encoder is pre-trained with unlabelled data in a self-supervised manner.
Moreover, a random masking mechanism is explicitly proposed to work with the Transformer feature encoder, which aims to improve the robustness of the feature encoder.
In addition, we have explored an ensemble-based network architecture in the downstream phase. Experimental results demonstrate that the designed ensemble network with random masking achieves top performance.
The findings of this research provide a new perspective and insights for cough classification.

\begin{acks}
This research is supported by Australian Research Council (ARC) Discovery Project \textit{DP190101485}. We would also like to thank the COVID-19 Sounds App team of the Department of Computer Science and Technology of the University of Cambridge for access to the COVID-19 sound dataset and Project Coswara by Indian Institute of Science (IISc) Bangalore for the Coswara dataset. 
\end{acks}

\bibliographystyle{ACM-Reference-Format}
\bibliography{main}


\begin{thebibliography}{33}


\ifx \showCODEN    \undefined \def \showCODEN     #1{\unskip}     \fi
\ifx \showDOI      \undefined \def \showDOI       #1{#1}\fi
\ifx \showISBNx    \undefined \def \showISBNx     #1{\unskip}     \fi
\ifx \showISBNxiii \undefined \def \showISBNxiii  #1{\unskip}     \fi
\ifx \showISSN     \undefined \def \showISSN      #1{\unskip}     \fi
\ifx \showLCCN     \undefined \def \showLCCN      #1{\unskip}     \fi
\ifx \shownote     \undefined \def \shownote      #1{#1}          \fi
\ifx \showarticletitle \undefined \def \showarticletitle #1{#1}   \fi
\ifx \showURL      \undefined \def \showURL       {\relax}        \fi
\providecommand\bibfield[2]{#2}
\providecommand\bibinfo[2]{#2}
\providecommand\natexlab[1]{#1}
\providecommand\showeprint[2][]{arXiv:#2}

\bibitem[\protect\citeauthoryear{Amoh and Odame}{Amoh and Odame}{2015}]%
        {amoh2015deepcough}
\bibfield{author}{\bibinfo{person}{Justice Amoh} {and} \bibinfo{person}{Kofi
  Odame}.} \bibinfo{year}{2015}\natexlab{}.
\newblock \showarticletitle{DeepCough: A deep convolutional neural network in a
  wearable cough detection system}. In \bibinfo{booktitle}{\emph{2015 IEEE
  Biomedical Circuits and Systems Conference (BioCAS)}}. IEEE,
  \bibinfo{pages}{1--4}.
\newblock


\bibitem[\protect\citeauthoryear{Amoh and Odame}{Amoh and Odame}{2016}]%
        {amoh2016deep}
\bibfield{author}{\bibinfo{person}{Justice Amoh} {and} \bibinfo{person}{Kofi
  Odame}.} \bibinfo{year}{2016}\natexlab{}.
\newblock \showarticletitle{Deep neural networks for identifying cough sounds}.
\newblock \bibinfo{journal}{\emph{IEEE transactions on biomedical circuits and
  systems}} \bibinfo{volume}{10}, \bibinfo{number}{5} (\bibinfo{year}{2016}),
  \bibinfo{pages}{1003--1011}.
\newblock


\bibitem[\protect\citeauthoryear{Bagad, Dalmia, Doshi, Nagrani, Bhamare,
  Mahale, Rane, Agarwal, and Panicker}{Bagad et~al\mbox{.}}{2020}]%
        {bagad2020cough}
\bibfield{author}{\bibinfo{person}{Piyush Bagad}, \bibinfo{person}{Aman
  Dalmia}, \bibinfo{person}{Jigar Doshi}, \bibinfo{person}{Arsha Nagrani},
  \bibinfo{person}{Parag Bhamare}, \bibinfo{person}{Amrita Mahale},
  \bibinfo{person}{Saurabh Rane}, \bibinfo{person}{Neeraj Agarwal}, {and}
  \bibinfo{person}{Rahul Panicker}.} \bibinfo{year}{2020}\natexlab{}.
\newblock \showarticletitle{Cough against covid: Evidence of covid-19 signature
  in cough sounds}.
\newblock \bibinfo{journal}{\emph{arXiv preprint arXiv:2009.08790}}
  (\bibinfo{year}{2020}).
\newblock


\bibitem[\protect\citeauthoryear{Bales, Nabeel, John, Masood, Qureshi, Farooq,
  Posokhova, and Imran}{Bales et~al\mbox{.}}{2020}]%
        {bales2020can}
\bibfield{author}{\bibinfo{person}{Charles Bales}, \bibinfo{person}{Muhammad
  Nabeel}, \bibinfo{person}{Charles~N John}, \bibinfo{person}{Usama Masood},
  \bibinfo{person}{Haneya~N Qureshi}, \bibinfo{person}{Hasan Farooq},
  \bibinfo{person}{Iryna Posokhova}, {and} \bibinfo{person}{Ali Imran}.}
  \bibinfo{year}{2020}\natexlab{}.
\newblock \showarticletitle{Can machine learning be used to recognize and
  diagnose coughs?}. In \bibinfo{booktitle}{\emph{2020 International Conference
  on e-Health and Bioengineering (EHB)}}. IEEE, \bibinfo{pages}{1--4}.
\newblock


\bibitem[\protect\citeauthoryear{Brown, Chauhan, Grammenos, Han,
  Hasthanasombat, Spathis, Xia, Cicuta, and Mascolo}{Brown
  et~al\mbox{.}}{2020}]%
        {brown2020exploring}
\bibfield{author}{\bibinfo{person}{Chlo{\"{e}} Brown},
  \bibinfo{person}{Jagmohan Chauhan}, \bibinfo{person}{Andreas Grammenos},
  \bibinfo{person}{Jing Han}, \bibinfo{person}{Apinan Hasthanasombat},
  \bibinfo{person}{Dimitris Spathis}, \bibinfo{person}{Tong Xia},
  \bibinfo{person}{Pietro Cicuta}, {and} \bibinfo{person}{Cecilia Mascolo}.}
  \bibinfo{year}{2020}\natexlab{}.
\newblock \showarticletitle{Exploring Automatic Diagnosis of {COVID-19} from
  Crowdsourced Respiratory Sound Data}. In \bibinfo{booktitle}{\emph{{KDD} '20:
  The 26th {ACM} {SIGKDD} Conference on Knowledge Discovery and Data Mining,
  Virtual Event, CA, USA, August 23-27, 2020}},
  \bibfield{editor}{\bibinfo{person}{Rajesh Gupta}, \bibinfo{person}{Yan Liu},
  \bibinfo{person}{Jiliang Tang}, {and} \bibinfo{person}{B.~Aditya Prakash}}
  (Eds.). \bibinfo{publisher}{{ACM}}, \bibinfo{pages}{3474--3484}.
\newblock
\urldef\tempurl%
\url{https://dl.acm.org/doi/10.1145/3394486.3412865}
\showURL{%
\tempurl}


\bibitem[\protect\citeauthoryear{Chatrzarrin, Arcelus, Goubran, and
  Knoefel}{Chatrzarrin et~al\mbox{.}}{2011}]%
        {chatrzarrin2011feature}
\bibfield{author}{\bibinfo{person}{Hanieh Chatrzarrin}, \bibinfo{person}{Amaya
  Arcelus}, \bibinfo{person}{Rafik Goubran}, {and} \bibinfo{person}{Frank
  Knoefel}.} \bibinfo{year}{2011}\natexlab{}.
\newblock \showarticletitle{Feature extraction for the differentiation of dry
  and wet cough sounds}. In \bibinfo{booktitle}{\emph{2011 IEEE international
  symposium on medical measurements and applications}}. IEEE,
  \bibinfo{pages}{162--166}.
\newblock


\bibitem[\protect\citeauthoryear{Chen, Kornblith, Norouzi, and Hinton}{Chen
  et~al\mbox{.}}{2020}]%
        {chen2020simple}
\bibfield{author}{\bibinfo{person}{Ting Chen}, \bibinfo{person}{Simon
  Kornblith}, \bibinfo{person}{Mohammad Norouzi}, {and}
  \bibinfo{person}{Geoffrey Hinton}.} \bibinfo{year}{2020}\natexlab{}.
\newblock \showarticletitle{A simple framework for contrastive learning of
  visual representations}. In \bibinfo{booktitle}{\emph{International
  conference on machine learning}}. PMLR, \bibinfo{pages}{1597--1607}.
\newblock


\bibitem[\protect\citeauthoryear{Chung, Gulcehre, Cho, and Bengio}{Chung
  et~al\mbox{.}}{2014}]%
        {chung2014empirical}
\bibfield{author}{\bibinfo{person}{Junyoung Chung}, \bibinfo{person}{Caglar
  Gulcehre}, \bibinfo{person}{KyungHyun Cho}, {and} \bibinfo{person}{Yoshua
  Bengio}.} \bibinfo{year}{2014}\natexlab{}.
\newblock \showarticletitle{{Empirical Evaluation of Gated Recurrent Neural
  Networks on Sequence Modeling}}.
\newblock \bibinfo{journal}{\emph{arXiv preprint arXiv:1412.3555}}
  (\bibinfo{year}{2014}).
\newblock


\bibitem[\protect\citeauthoryear{Devlin, Chang, Lee, and Toutanova}{Devlin
  et~al\mbox{.}}{2019}]%
        {bert}
\bibfield{author}{\bibinfo{person}{Jacob Devlin}, \bibinfo{person}{Ming{-}Wei
  Chang}, \bibinfo{person}{Kenton Lee}, {and} \bibinfo{person}{Kristina
  Toutanova}.} \bibinfo{year}{2019}\natexlab{}.
\newblock \showarticletitle{{BERT:} Pre-training of Deep Bidirectional
  Transformers for Language Understanding}. In
  \bibinfo{booktitle}{\emph{Proceedings of the Conference of the North American
  Chapter of the Association for Computational Linguistics: Human Language
  Technologies,}}, \bibfield{editor}{\bibinfo{person}{Jill Burstein},
  \bibinfo{person}{Christy Doran}, {and} \bibinfo{person}{Thamar Solorio}}
  (Eds.). \bibinfo{pages}{4171--4186}.
\newblock


\bibitem[\protect\citeauthoryear{Gemmeke, Ellis, Freedman, Jansen, Lawrence,
  Moore, Plakal, and Ritter}{Gemmeke et~al\mbox{.}}{2017}]%
        {gemmeke2017audio}
\bibfield{author}{\bibinfo{person}{Jort~F Gemmeke}, \bibinfo{person}{Daniel~PW
  Ellis}, \bibinfo{person}{Dylan Freedman}, \bibinfo{person}{Aren Jansen},
  \bibinfo{person}{Wade Lawrence}, \bibinfo{person}{R~Channing Moore},
  \bibinfo{person}{Manoj Plakal}, {and} \bibinfo{person}{Marvin Ritter}.}
  \bibinfo{year}{2017}\natexlab{}.
\newblock \showarticletitle{Audio set: An ontology and human-labeled dataset
  for audio events}. In \bibinfo{booktitle}{\emph{2017 IEEE International
  Conference on Acoustics, Speech and Signal Processing (ICASSP)}}. IEEE,
  \bibinfo{pages}{776--780}.
\newblock


\bibitem[\protect\citeauthoryear{Han, Chan, Choy, and Pun}{Han
  et~al\mbox{.}}{2006}]%
        {han2006efficient}
\bibfield{author}{\bibinfo{person}{Wei Han}, \bibinfo{person}{Cheong-Fat Chan},
  \bibinfo{person}{Chiu-Sing Choy}, {and} \bibinfo{person}{Kong-Pang Pun}.}
  \bibinfo{year}{2006}\natexlab{}.
\newblock \showarticletitle{An efficient MFCC extraction method in speech
  recognition}. In \bibinfo{booktitle}{\emph{2006 IEEE international symposium
  on circuits and systems}}. IEEE, \bibinfo{pages}{4--pp}.
\newblock


\bibitem[\protect\citeauthoryear{Hassan, Shahin, and Alsabek}{Hassan
  et~al\mbox{.}}{2020}]%
        {hassan2020covid}
\bibfield{author}{\bibinfo{person}{Abdelfatah Hassan}, \bibinfo{person}{Ismail
  Shahin}, {and} \bibinfo{person}{Mohamed~Bader Alsabek}.}
  \bibinfo{year}{2020}\natexlab{}.
\newblock \showarticletitle{Covid-19 detection system using recurrent neural
  networks}. In \bibinfo{booktitle}{\emph{2020 International Conference on
  Communications, Computing, Cybersecurity, and Informatics (CCCI)}}. IEEE,
  \bibinfo{pages}{1--5}.
\newblock


\bibitem[\protect\citeauthoryear{He, Fan, Wu, Xie, and Girshick}{He
  et~al\mbox{.}}{2020}]%
        {he2020momentum}
\bibfield{author}{\bibinfo{person}{Kaiming He}, \bibinfo{person}{Haoqi Fan},
  \bibinfo{person}{Yuxin Wu}, \bibinfo{person}{Saining Xie}, {and}
  \bibinfo{person}{Ross Girshick}.} \bibinfo{year}{2020}\natexlab{}.
\newblock \showarticletitle{Momentum contrast for unsupervised visual
  representation learning}. In \bibinfo{booktitle}{\emph{Proceedings of the
  IEEE/CVF Conference on Computer Vision and Pattern Recognition}}.
  \bibinfo{pages}{9729--9738}.
\newblock


\bibitem[\protect\citeauthoryear{He, Zhang, Ren, and Sun}{He
  et~al\mbox{.}}{2016}]%
        {he2016deep}
\bibfield{author}{\bibinfo{person}{Kaiming He}, \bibinfo{person}{Xiangyu
  Zhang}, \bibinfo{person}{Shaoqing Ren}, {and} \bibinfo{person}{Jian Sun}.}
  \bibinfo{year}{2016}\natexlab{}.
\newblock \showarticletitle{Deep residual learning for image recognition}. In
  \bibinfo{booktitle}{\emph{Proceedings of the IEEE conference on computer
  vision and pattern recognition}}. \bibinfo{pages}{770--778}.
\newblock


\bibitem[\protect\citeauthoryear{Hershey, Chaudhuri, Ellis, Gemmeke, Jansen,
  Moore, Plakal, Platt, Saurous, Seybold, et~al\mbox{.}}{Hershey
  et~al\mbox{.}}{2017}]%
        {hershey2017cnn}
\bibfield{author}{\bibinfo{person}{Shawn Hershey}, \bibinfo{person}{Sourish
  Chaudhuri}, \bibinfo{person}{Daniel~PW Ellis}, \bibinfo{person}{Jort~F
  Gemmeke}, \bibinfo{person}{Aren Jansen}, \bibinfo{person}{R~Channing Moore},
  \bibinfo{person}{Manoj Plakal}, \bibinfo{person}{Devin Platt},
  \bibinfo{person}{Rif~A Saurous}, \bibinfo{person}{Bryan Seybold},
  {et~al\mbox{.}}} \bibinfo{year}{2017}\natexlab{}.
\newblock \showarticletitle{CNN architectures for large-scale audio
  classification}. In \bibinfo{booktitle}{\emph{2017 ieee international
  conference on acoustics, speech and signal processing (icassp)}}. IEEE,
  \bibinfo{pages}{131--135}.
\newblock


\bibitem[\protect\citeauthoryear{Hochreiter and Schmidhuber}{Hochreiter and
  Schmidhuber}{1997}]%
        {hochreiter1997long}
\bibfield{author}{\bibinfo{person}{Sepp Hochreiter} {and}
  \bibinfo{person}{J{\"u}rgen Schmidhuber}.} \bibinfo{year}{1997}\natexlab{}.
\newblock \showarticletitle{Long Short-term Memory}.
\newblock \bibinfo{journal}{\emph{Neural computation}} \bibinfo{volume}{9},
  \bibinfo{number}{8} (\bibinfo{year}{1997}), \bibinfo{pages}{1735--1780}.
\newblock


\bibitem[\protect\citeauthoryear{Imran, Posokhova, Qureshi, Masood, Riaz, Ali,
  John, Hussain, and Nabeel}{Imran et~al\mbox{.}}{2020}]%
        {imran2020ai4covid}
\bibfield{author}{\bibinfo{person}{Ali Imran}, \bibinfo{person}{Iryna
  Posokhova}, \bibinfo{person}{Haneya~N Qureshi}, \bibinfo{person}{Usama
  Masood}, \bibinfo{person}{Muhammad~Sajid Riaz}, \bibinfo{person}{Kamran Ali},
  \bibinfo{person}{Charles~N John}, \bibinfo{person}{MD~Iftikhar Hussain},
  {and} \bibinfo{person}{Muhammad Nabeel}.} \bibinfo{year}{2020}\natexlab{}.
\newblock \showarticletitle{AI4COVID-19: AI enabled preliminary diagnosis for
  COVID-19 from cough samples via an app}.
\newblock \bibinfo{journal}{\emph{Informatics in Medicine Unlocked}}
  \bibinfo{volume}{20} (\bibinfo{year}{2020}), \bibinfo{pages}{100378}.
\newblock


\bibitem[\protect\citeauthoryear{Jiang, Li, Cao, Zhang, Zou, Han, and Li}{Jiang
  et~al\mbox{.}}{2020}]%
        {jiang2020speech}
\bibfield{author}{\bibinfo{person}{Dongwei Jiang}, \bibinfo{person}{Wubo Li},
  \bibinfo{person}{Miao Cao}, \bibinfo{person}{Ruixiong Zhang},
  \bibinfo{person}{Wei Zou}, \bibinfo{person}{Kun Han}, {and}
  \bibinfo{person}{Xiangang Li}.} \bibinfo{year}{2020}\natexlab{}.
\newblock \showarticletitle{Speech SIMCLR: Combining Contrastive and
  Reconstruction Objective for Self-supervised Speech Representation Learning}.
\newblock \bibinfo{journal}{\emph{arXiv preprint arXiv:2010.13991}}
  (\bibinfo{year}{2020}).
\newblock


\bibitem[\protect\citeauthoryear{Kingma and Ba}{Kingma and Ba}{2015}]%
        {kingma2014adam}
\bibfield{author}{\bibinfo{person}{Diederik~P. Kingma} {and}
  \bibinfo{person}{Jimmy Ba}.} \bibinfo{year}{2015}\natexlab{}.
\newblock \showarticletitle{Adam: {A} Method for Stochastic Optimization}. In
  \bibinfo{booktitle}{\emph{3rd International Conference on Learning
  Representations, {ICLR} 2015, San Diego, CA, USA, May 7-9, 2015, Conference
  Track Proceedings}}, \bibfield{editor}{\bibinfo{person}{Yoshua Bengio} {and}
  \bibinfo{person}{Yann LeCun}} (Eds.).
\newblock


\bibitem[\protect\citeauthoryear{Li, Lin, Tsai, Yang, and Lin}{Li
  et~al\mbox{.}}{2017}]%
        {li2017design}
\bibfield{author}{\bibinfo{person}{Shih-Hong Li}, \bibinfo{person}{Bor-Shing
  Lin}, \bibinfo{person}{Chen-Han Tsai}, \bibinfo{person}{Cheng-Ta Yang}, {and}
  \bibinfo{person}{Bor-Shyh Lin}.} \bibinfo{year}{2017}\natexlab{}.
\newblock \showarticletitle{Design of wearable breathing sound monitoring
  system for real-time wheeze detection}.
\newblock \bibinfo{journal}{\emph{Sensors}} \bibinfo{volume}{17},
  \bibinfo{number}{1} (\bibinfo{year}{2017}), \bibinfo{pages}{171}.
\newblock


\bibitem[\protect\citeauthoryear{Mouawad, Dubnov, and Dubnov}{Mouawad
  et~al\mbox{.}}{2021}]%
        {mouawad2021robust}
\bibfield{author}{\bibinfo{person}{Pauline Mouawad}, \bibinfo{person}{Tammuz
  Dubnov}, {and} \bibinfo{person}{Shlomo Dubnov}.}
  \bibinfo{year}{2021}\natexlab{}.
\newblock \showarticletitle{Robust Detection of COVID-19 in Cough Sounds: Using
  Recurrence Dynamics and Variable Markov Model}.
\newblock \bibinfo{journal}{\emph{Sn Computer Science}} \bibinfo{volume}{2},
  \bibinfo{number}{1} (\bibinfo{year}{2021}).
\newblock


\bibitem[\protect\citeauthoryear{Oletic and Bilas}{Oletic and Bilas}{2016}]%
        {oletic2016energy}
\bibfield{author}{\bibinfo{person}{Dinko Oletic} {and} \bibinfo{person}{Vedran
  Bilas}.} \bibinfo{year}{2016}\natexlab{}.
\newblock \showarticletitle{Energy-efficient respiratory sounds sensing for
  personal mobile asthma monitoring}.
\newblock \bibinfo{journal}{\emph{Ieee sensors journal}} \bibinfo{volume}{16},
  \bibinfo{number}{23} (\bibinfo{year}{2016}), \bibinfo{pages}{8295--8303}.
\newblock


\bibitem[\protect\citeauthoryear{Oord, Li, and Vinyals}{Oord
  et~al\mbox{.}}{2018}]%
        {oord2018representation}
\bibfield{author}{\bibinfo{person}{Aaron van~den Oord}, \bibinfo{person}{Yazhe
  Li}, {and} \bibinfo{person}{Oriol Vinyals}.} \bibinfo{year}{2018}\natexlab{}.
\newblock \showarticletitle{Representation learning with contrastive predictive
  coding}.
\newblock \bibinfo{journal}{\emph{arXiv preprint arXiv:1807.03748}}
  (\bibinfo{year}{2018}).
\newblock


\bibitem[\protect\citeauthoryear{Orlandic, Teijeiro, and Atienza}{Orlandic
  et~al\mbox{.}}{2020}]%
        {orlandic2020coughvid}
\bibfield{author}{\bibinfo{person}{Lara Orlandic}, \bibinfo{person}{Tomas
  Teijeiro}, {and} \bibinfo{person}{David Atienza}.}
  \bibinfo{year}{2020}\natexlab{}.
\newblock \showarticletitle{The COUGHVID crowdsourcing dataset: A corpus for
  the study of large-scale cough analysis algorithms}.
\newblock \bibinfo{journal}{\emph{arXiv preprint arXiv:2009.11644}}
  (\bibinfo{year}{2020}).
\newblock


\bibitem[\protect\citeauthoryear{Pahar, Klopper, Warren, and Niesler}{Pahar
  et~al\mbox{.}}{2020}]%
        {pahar2020covid}
\bibfield{author}{\bibinfo{person}{Madhurananda Pahar}, \bibinfo{person}{Marisa
  Klopper}, \bibinfo{person}{Robin Warren}, {and} \bibinfo{person}{Thomas
  Niesler}.} \bibinfo{year}{2020}\natexlab{}.
\newblock \showarticletitle{COVID-19 Cough Classification using Machine
  Learning and Global Smartphone Recordings}.
\newblock \bibinfo{journal}{\emph{arXiv preprint arXiv:2012.01926}}
  (\bibinfo{year}{2020}).
\newblock


\bibitem[\protect\citeauthoryear{Saeed, Grangier, and Zeghidour}{Saeed
  et~al\mbox{.}}{2020}]%
        {saeed2020contrastive}
\bibfield{author}{\bibinfo{person}{Aaqib Saeed}, \bibinfo{person}{David
  Grangier}, {and} \bibinfo{person}{Neil Zeghidour}.}
  \bibinfo{year}{2020}\natexlab{}.
\newblock \showarticletitle{Contrastive Learning of General-Purpose Audio
  Representations}.
\newblock \bibinfo{journal}{\emph{arXiv preprint arXiv:2010.10915}}
  (\bibinfo{year}{2020}).
\newblock


\bibitem[\protect\citeauthoryear{Schuller, Coppock, and Gaskell}{Schuller
  et~al\mbox{.}}{2020}]%
        {schuller2020detecting}
\bibfield{author}{\bibinfo{person}{Bj{\"o}rn~W Schuller},
  \bibinfo{person}{Harry Coppock}, {and} \bibinfo{person}{Alexander Gaskell}.}
  \bibinfo{year}{2020}\natexlab{}.
\newblock \showarticletitle{Detecting COVID-19 from Breathing and Coughing
  Sounds using Deep Neural Networks}.
\newblock \bibinfo{journal}{\emph{arXiv preprint arXiv:2012.14553}}
  (\bibinfo{year}{2020}).
\newblock


\bibitem[\protect\citeauthoryear{Sharma, Krishnan, Kumar, Ramoji, Chetupalli,
  Nirmala, Kumar~Ghosh, and Ganapathy}{Sharma et~al\mbox{.}}{2020}]%
        {sharma2020coswara}
\bibfield{author}{\bibinfo{person}{N Sharma}, \bibinfo{person}{P Krishnan},
  \bibinfo{person}{R Kumar}, \bibinfo{person}{S Ramoji}, \bibinfo{person}{SR
  Chetupalli}, \bibinfo{person}{R Nirmala}, \bibinfo{person}{P Kumar~Ghosh},
  {and} \bibinfo{person}{S Ganapathy}.} \bibinfo{year}{2020}\natexlab{}.
\newblock \showarticletitle{Coswara-A database of breathing, cough, and voice
  sounds for COVID-19 diagnosis}. In \bibinfo{booktitle}{\emph{Proceedings of
  the Annual Conference of the International Speech Communication Association,
  INTERSPEECH}}, Vol.~\bibinfo{volume}{2020}. International Speech
  Communication Association, \bibinfo{pages}{4811--4815}.
\newblock


\bibitem[\protect\citeauthoryear{Simonyan and Zisserman}{Simonyan and
  Zisserman}{2014}]%
        {simonyan2014very}
\bibfield{author}{\bibinfo{person}{Karen Simonyan} {and}
  \bibinfo{person}{Andrew Zisserman}.} \bibinfo{year}{2014}\natexlab{}.
\newblock \showarticletitle{Very deep convolutional networks for large-scale
  image recognition}.
\newblock \bibinfo{journal}{\emph{arXiv preprint arXiv:1409.1556}}
  (\bibinfo{year}{2014}).
\newblock


\bibitem[\protect\citeauthoryear{Tan and Le}{Tan and Le}{2019}]%
        {tan2019efficientnet}
\bibfield{author}{\bibinfo{person}{Mingxing Tan} {and} \bibinfo{person}{Quoc
  Le}.} \bibinfo{year}{2019}\natexlab{}.
\newblock \showarticletitle{Efficientnet: Rethinking model scaling for
  convolutional neural networks}. In \bibinfo{booktitle}{\emph{International
  Conference on Machine Learning}}. PMLR, \bibinfo{pages}{6105--6114}.
\newblock


\bibitem[\protect\citeauthoryear{Vaswani, Shazeer, Parmar, Uszkoreit, Jones,
  Gomez, Kaiser, and Polosukhin}{Vaswani et~al\mbox{.}}{2017}]%
        {vaswani2017attention}
\bibfield{author}{\bibinfo{person}{Ashish Vaswani}, \bibinfo{person}{Noam
  Shazeer}, \bibinfo{person}{Niki Parmar}, \bibinfo{person}{Jakob Uszkoreit},
  \bibinfo{person}{Llion Jones}, \bibinfo{person}{Aidan~N Gomez},
  \bibinfo{person}{{\L}ukasz Kaiser}, {and} \bibinfo{person}{Illia
  Polosukhin}.} \bibinfo{year}{2017}\natexlab{}.
\newblock \showarticletitle{Attention is all you need}. In
  \bibinfo{booktitle}{\emph{Proceedings of the 31st International Conference on
  Neural Information Processing Systems}}. \bibinfo{pages}{6000--6010}.
\newblock


\bibitem[\protect\citeauthoryear{Wu, Huang, Zhang, and Chawla}{Wu
  et~al\mbox{.}}{2020}]%
        {wu2020hierarchically}
\bibfield{author}{\bibinfo{person}{Xian Wu}, \bibinfo{person}{Chao Huang},
  \bibinfo{person}{Chuxu Zhang}, {and} \bibinfo{person}{Nitesh~V Chawla}.}
  \bibinfo{year}{2020}\natexlab{}.
\newblock \showarticletitle{Hierarchically structured transformer networks for
  fine-grained spatial event forecasting}. In
  \bibinfo{booktitle}{\emph{Proceedings of The Web Conference 2020}}.
  \bibinfo{pages}{2320--2330}.
\newblock


\bibitem[\protect\citeauthoryear{Xue and Salim}{Xue and Salim}{2021}]%
        {xue2020trailer}
\bibfield{author}{\bibinfo{person}{Hao Xue} {and} \bibinfo{person}{Flora~D
  Salim}.} \bibinfo{year}{2021}\natexlab{}.
\newblock \showarticletitle{TERMCast: Temporal Relation Modeling for Effective
  Urban Flow Forecasting}. In \bibinfo{booktitle}{\emph{Pacific-Asia Conference
  on Knowledge Discovery and Data Mining}}.
\newblock


\end{thebibliography}

\end{document}